\documentstyle[11pt,psfig,twoside]{article}

\begin{document}

\newcommand{\dd}{\,{\rm d}}
\newcommand{\ie}{{\it i.e.},\,}
\newcommand{\etal}{{\it et al.\ }}
\newcommand{\eg}{{\it e.g.},\,}
\newcommand{\cf}{{\it cf.\ }}
\newcommand{\vs}{{\it vs.\ }}
\newcommand{\zdot}{\makebox[0pt][l]{.}}
\newcommand{\up}[1]{\ifmmode^{\rm #1}\else$^{\rm #1}$\fi}
\newcommand{\dn}[1]{\ifmmode_{\rm #1}\else$_{\rm #1}$\fi}
\newcommand{\upd}{\up{d}}
\newcommand{\uph}{\up{h}}
\newcommand{\upm}{\up{m}}
\newcommand{\ups}{\up{s}}
\newcommand{\arcd}{\ifmmode^{\circ}\else$^{\circ}$\fi}
\newcommand{\arcm}{\ifmmode{'}\else$'$\fi}
\newcommand{\arcs}{\ifmmode{''}\else$''$\fi}
\newcommand{\MS}{{\rm M}\ifmmode_{\odot}\else$_{\odot}$\fi}
\newcommand{\RS}{{\rm R}\ifmmode_{\odot}\else$_{\odot}$\fi}
\newcommand{\LS}{{\rm L}\ifmmode_{\odot}\else$_{\odot}$\fi}

\newcommand{\Abstract}[2]{{\footnotesize\begin{center}ABSTRACT\end{center}
\vspace{1mm}\par#1\par
\noindent
{~}{\it #2}}}

\newcommand{\TabCap}[2]{\begin{center}\parbox[t]{#1}{\begin{center}
  \small {\spaceskip 2pt plus 1pt minus 1pt T a b l e}
  \refstepcounter{table}\thetable \\[2mm]
  \footnotesize #2 \end{center}}\end{center}}

\newcommand{\TableSep}[2]{\begin{table}[p]\vspace{#1}
\TabCap{#2}\end{table}}

\newcommand{\FigCap}[1]{\footnotesize\par\noindent Fig.\  %
  \refstepcounter{figure}\thefigure. #1\par}

\newcommand{\TableFont}{\footnotesize}
\newcommand{\TableFontIt}{\ttit}
\newcommand{\SetTableFont}[1]{\renewcommand{\TableFont}{#1}}

\newcommand{\MakeTable}[4]{\begin{table}[htb]\TabCap{#2}{#3}
  \begin{center} \TableFont \begin{tabular}{#1} #4 
  \end{tabular}\end{center}\end{table}}

\newcommand{\MakeTableSep}[4]{\begin{table}[p]\TabCap{#2}{#3}
  \begin{center} \TableFont \begin{tabular}{#1} #4 
  \end{tabular}\end{center}\end{table}}

\newenvironment{references}%
{
\footnotesize \frenchspacing
\renewcommand{\thesection}{}
\renewcommand{\in}{{\rm in }}
\renewcommand{\AA}{Astron.\ Astrophys.}
\newcommand{\AAS}{Astron.~Astrophys.~Suppl.~Ser.}
\newcommand{\ApJ}{Astrophys.\ J.}
\newcommand{\ApJS}{Astrophys.\ J.~Suppl.~Ser.}
\newcommand{\ApJL}{Astrophys.\ J.~Letters}
\newcommand{\AJ}{Astron.\ J.}
\newcommand{\IBVS}{IBVS}
\newcommand{\PASP}{P.A.S.P.}
\newcommand{\Acta}{Acta Astron.}
\newcommand{\MNRAS}{MNRAS}
\renewcommand{\and}{{\rm and }}
\section{{\rm REFERENCES}}
\sloppy \hyphenpenalty10000
\begin{list}{}{\leftmargin1cm\listparindent-1cm
\itemindent\listparindent\parsep0pt\itemsep0pt}}%
{\end{list}\vspace{2mm}}

\def\TYLDA{~}
\newlength{\DW}
\settowidth{\DW}{0}
\newcommand{\dw}{\hspace{\DW}}

\newcommand{\refitem}[5]{\item[]{#1} #2%
\def\REFARG{#3}\ifx\REFARG\TYLDA\else, {\it#3}\fi
\def\REFARG{#4}\ifx\REFARG\TYLDA\else, {\bf#4}\fi
\def\REFARG{#5}\ifx\REFARG\TYLDA\else, {#5}\fi.}

\newcommand{\Section}[1]{\section{#1}}
\newcommand{\Subsection}[1]{\subsection{#1}}
\newcommand{\Acknow}[1]{\par\vspace{5mm}{\bf Acknowledgements.} #1}
\pagestyle{myheadings}

\def\thefootnote{\fnsymbol{footnote}}
\begin{center}
{\large\bf The Optical Gravitational Lensing Experiment.\\
%\vskip3pt
Cepheids in the Magellanic Clouds.\\
%\vskip3pt
III.  Period-Luminosity-Color and Period-Luminosity Relations
%\vskip3pt
of Classical Cepheids\footnote{Based on  observations obtained
with the 1.3~m Warsaw telescope at the Las Campanas  Observatory of the
Carnegie Institution of Washington.}}
\vskip0.8cm
{\bf
A.~~U~d~a~l~s~k~i$^1$,~~M.~~S~z~y~m~a~{\'n}~s~k~i$^1$,~~M.~~K~u~b~i~a~k$^1$,
~~G.~~P~i~e~t~r~z~y~\'n~s~k~i$^1$,~~I.~~S~o~s~z~y~{\'n}~s~k~i$^1$,
~~P.~~W~o~\'z~n~i~a~k$^2$,~~ and~~K.~~\.Z~e~b~r~u~\'n$^1$}
\vskip3mm
{$^1$Warsaw University Observatory, Al.~Ujazdowskie~4, 00-478~Warszawa, Poland\\
e-mail: (udalski,msz,mk,pietrzyn,soszynsk,zebrun)@sirius.astrouw.edu.pl\\
$^2$ Princeton University Observatory, Princeton, NJ 08544-1001, USA\\
e-mail: wozniak@astro.princeton.edu}
\end{center}

\Abstract{

We present Period-Luminosity-Color and Period-Luminosity relations of
classical Ce\-pheids constructed for about 1280 Cepheids from the LMC
and 2140  from the SMC. High quality {\it BVI} observations (120--360
epochs in the {\it I}-band and 15--40 in the {\it BV}-bands) were
collected during the OGLE-II microlensing experiment. The {\it I}-band
diagrams of the LMC show very small scatter, $\sigma=0.074$~mag,
indicating that Cepheid variables can potentially be a very good
standard candle.

We compare relations of fundamental mode Cepheids from the LMC and SMC
and we do not find significant differences of slopes of the
Period-Luminosity-Color and Period-Luminosity relations in these
galaxies. For the first overtone Cepheids a small change of the slope of
Period-Luminosity relation is possible. 

We determine the difference of distance moduli between the SMC and LMC
with Cepheid relations and compare the result with the difference obtained
with other standard candles: RR~Lyr and red clump stars. Results are
very consistent and indicate that the values of zero points of the
fundamental mode Cepheid relations are similar in these galaxies. The
mean difference of distance moduli between the SMC and LMC is equal to
$\mu_{SMC}-\mu_{LMC}=0.51\pm0.03$~mag.

We calibrate the Period-Luminosity-Color and  Period-Luminosity
relations for classical, fundamental mode  Cepheids using the observed
LMC relations and adopting the short LMC distance modulus,
$\mu_{LMC}=18.22\pm0.05$~mag, resulting from the recent determinations with
eclipsing system HV2274, RR~Lyr and  red clump stars and consistent with
observations of Cepheids in NGC4258 galaxy, to which precise geometric
distance is known.

Finally, we determine a constraint on the absolute magnitude of Cepheids
by comparison of their mean {\it V}-band magnitude with that of RR~Lyr
stars in both Magellanic Clouds. The 10-day period, fundamental mode
Cepheid is on average $4.63\pm0.05$~mag brighter than RR~Lyr stars of
LMC metallicity which with the most likely calibration of the brightness
of RR~Lyr stars yields $M_V^{C,10}=-3.92\pm0.09$~mag.}{~}

\Section{Introduction}

The Cepheid variable stars are among the most important objects of the
modern astrophysics. These relatively well understood pulsating stars
provide many empirical information on stellar structure, evolution etc.
But what makes them particularly attractive is the well known
correlation of their brightness with period, discovered yet at the
beginning of the 20th century (Leavitt 1912). This feature of Cepheids
and their large absolute brightness make them potentially ideal standard
candle for precise distance determination to extragalactic objects where
Cepheids have been discovered. However, after almost a century since its
discovery the calibration of $P-L$ relation  is still a subject of
considerable dispute.

While determination of precise periods of Cepheids is straightforward,
determination and calibration of Cepheid brightness is difficult. The
Galactic Cepheids are located so far from the Sun that the distance
determination to them can only be obtained by indirect, often very
uncertain methods. Even the Hipparcos satellite did not provide much
progress in this field as it measured parallaxes of only a few Galactic
Cepheids with accuracy better than 30\% (Feast and Catchpole  1997).
Galactic Cepheids are also usually highly reddened and accurate
determination of their brightness is not easy. The Magellanic Clouds,
where the $P-L$ relation was discovered, play an important role in this
field. Both the Large and Small Magellanic Clouds are known to contain
large population of Cepheids, located at approximately the same
distance. Therefore the slope of the $P-L$ relation of Cepheids was
usually derived based on the Magellanic Cloud objects. 

The Magellanic Cloud calibrations (Caldwell and Coulson 1986, Madore and
Freedman 1991, Laney and Stobie 1994) are based, however, on limited
samples of stars with photometry obtained many years ago mostly with
photoelectric and photographic techniques what in crowded fields may
lead to systematic uncertainties.  Unfortunately both galaxies which are
the best objects for studying properties of Cepheids have been very
rarely observed with modern CCD techniques. Situation has changed
dramatically in 1990s when large microlensing surveys started regular
observations of the Magellanic Clouds. Photometry of millions stars is a
natural by-product of these surveys, and for the first time precise
light curves of thousands of Cepheids could be obtained. The MACHO team
(Alcock \etal 1995) presented impressive $P-L$ diagram of Cepheids in
the LMC showing for the first time  clear division between the $P-L$
relation of the fundamental mode and first overtone Cepheids, merged in
previous data. Also the EROS group analyzed $P-L$ diagrams in the SMC and
LMC (Sasselov \etal 1997) suggesting dependence of the zero point of 
$P-L$ relation on metallicity. They also found a change of slope of the
short period fundamental mode Cepheids in the SMC (Bauer \etal 1999).

Both the MACHO and EROS data are taken in non-standard bands, therefore
they are not suitable for general $P-L$ relation calibrations. The
problem of good calibration of the $P-L$ relation becomes very urgent
now, because the Cepheid variables are routinely discovered  in many
galaxies by HST. The main goal of the HST Key Project (Kennicutt,
Freedman and Mould 1995) is determination of the Hubble constant. The
group uses, however, the universal $P-L$ relation based on very small
number of LMC Cepheids (Madore and Freedman 1991),  neglecting possible
population effects, and assuming the distance modulus to the LMC
$\mu_{LMC}=18.50$~mag. Any uncertainties of  calibration of the $P-L$
relation and the LMC distance propagate to any Cepheid based distance
and as a result to the value of the Hubble constant with all
astrophysical consequences as, for example, age of the Universe.

The Magellanic Clouds were added to the targets of the OGLE microlensing
search at the beginning of the second phase of the project -- OGLE-II
(Udalski, Kubiak and Szyma\'nski 1997) in January 1997. Observations are
collected in the standard {\it BVI}-bands and after more than two years
of observations the photometric databases consist of a few hundreds
epochs of a few millions stars from the LMC and SMC. The OGLE-II databases
were already searched for variable stars and Cepheids are very numerous
among them. Some results on Cepheid variables detected in the OGLE-II
data -- double-mode Cepheids and second overtone Cepheids in the SMC --
were already presented in previous papers of this series (Udalski \etal
1999a,b). Catalogs with {\it BVI} photometry of about 1400 Cepheids from
the LMC and 2300 from the SMC  will be released in the following papers.

In this paper we present analysis of the period-luminosity-color
($P-L-C$) and $P-L$ relations for Cepheids from the LMC and SMC. We
analyze mainly the fundamental mode pulsators because they are used as
distance indicators. We compare relations of both the LMC and SMC and
find that the slopes of relations in both galaxies are within errors
similar. The difference of distance moduli of the SMC and LMC inferred
with the Cepheid relations is consistent with that found with other
reliable standard candles indicating no dependence of the zero points of
Cepheid relations on population differences between the Clouds. We
calibrate the $P-L-C$ and $P-L$ relations using the short distance
modulus to the LMC ($\mu_{LMC}=18.22$~mag) resulting from the recent
determinations with other reliable distance indicators. Finally, we
compare {\it  V}-band magnitudes of Cepheids with that of RR~Lyr stars
providing a tight constraint on their absolute magnitude.

\Section{Observational Data}

The observational data presented in this paper were collected during the
second phase of the OGLE microlensing search with the 1.3-m Warsaw
telescope at the Las Campanas Observatory, Chile which is operated by
the Carnegie Institution of Washington.  The telescope was equipped with
the "first generation" camera with   a SITe ${2048\times2048}$ CCD
detector working in  drift-scan mode. The  pixel size was 24~$\mu$m
giving the 0.417 arcsec/pixel scale. Observations were performed in the
"slow" reading mode of CCD detector with the  gain 3.8~e$^-$/ADU and
readout noise of about 5.4~e$^-$. Details of the  instrumentation setup
can be found in Udalski, Kubiak and Szyma{\'n}ski  (1997).

Observations covered significant part of the central regions of both the
LMC and SMC. Practically the entire bars of these galaxies -- more than
4.5 square degrees (21 $14.2\times57$ arcmin driftscan fields) and about
2.4 square degrees (11 fields) were monitored regularly from January
1997 through June 1999 and June 1997 through March 1999 for the LMC and
SMC, respectively. Collected {\it BVI} data were reduced and calibrated
to the standard system. Accuracy of transformation to the standard
system was about $0.01-0.02$~mag. The photometric data of the SMC were
used to construct the {\it BVI} photometric  maps of the SMC (Udalski
\etal 1998b). The reader is referred to that paper for  more details
about methods of data reduction, tests on quality of photometric  data,
astrometry, location of observed fields etc. Quality of the LMC data is
similar and it will be fully described with release of the {\it BVI}
photometric  maps of the LMC in the near future.

OGLE-II photometric databases were already searched for variable stars,
in particular pulsating variables. About 1400 Cepheids were detected in
the LMC fields and 2300 in the SMC. Part of the SMC  sample, namely
double-mode Cepheids and second overtone Cepheids, was already presented
in Udalski \etal (1999a, 1999b). The LMC and SMC Cepheid catalogs will
be presented in the following papers of this series. They will describe
in detail methods of selection of Cepheid variables, determination of
mean photometry, completeness of the sample etc. 

In short, the light curves of each object consist of about 120--360
epochs in the {\it I}-band and about 15--40 in the {\it V} and {\it
B}-bands.  {\it BVI} photometry was available for all 11 driftscan
fields in the SMC. {\it B}-band photometry for the LMC is at the
moment of writing this paper less complete -- reductions of only 40\% of
fields were finished. For the remaining fields only {\it VI} photometry
was available. {\it B}-band photometry of these fields will be completed
after the next observing season.

The observational data used for construction of the $P-L-C$ and $P-L$
diagrams consist of {\it BVI} photometry and period of light variations.
The intensity-mean brightness of each object was derived by fitting the
light curve with fifth order Fourier series. Accuracy of the {\it
I}-band mean magnitudes is about a few thousands of magnitude. Accuracy
of the {\it BV}-band magnitudes is somewhat worse because of worse
sampling of light curves -- about 0.01~mag. Accuracy of periods is about
${7\cdot10^{-5}P}$.

\Section{Determination of the $P-L-C$ and $P-L$ Relations}

To derive the $P-L-C$ and $P-L$ relations for Cepheids in the LMC
and SMC, we selected all objects from the OGLE Catalogs of Cepheids
cataloged as fundamental mode (FU) and first overtone (FO) pulsators. In
the next step we corrected the mean brightness for interstellar
extinction. We used red clump giants as the reference brightness for
determination of the mean interstellar extinction. Red clump stars
are very numerous in both the LMC and SMC, their {\it I}-band magnitude
was shown to be independent on age of these stars in the wide range
of $2-10$~Gyr, and it is only slightly dependent on metallicity (Udalski
1998a,b). The latter correction is not important in this case because of
practically homogeneous environment of field stars in the LMC (Bica
\etal 1998) or SMC. Thus the mean brightness of red clump stars can
be a very good reference of brightness for monitoring extinction.
Similar method was used by Stanek (1996) for determination of extinction
map of Baade's Window in the Galactic bulge.

The reddening was determined in 84 lines-of-sight in the LMC and 11 in
the SMC. The total mean reddening of the observed fields was found to be
$E(B-V)=0.137$ and $E(B-V)=0.087$ in the LMC and SMC, respectively.  Its
fluctuations, more significant in the LMC, and non-uniform distribution
of Cepheids resulted in somewhat larger  total mean reddening of these
stars: $E(B-V)=0.147$ and $E(B-V)=0.092$ for the LMC and SMC Cepheid
samples, respectively.  More details on extinction determination will be
provided with the release of Catalogs of Cepheids from the LMC and SMC.

It is obvious that our extinction corrections remove effects of
extinction in statistical sense only but because of huge sample of
Cepheids presented in this paper our approach is well justified.  To
test whether our extinction correction indeed removes inhomogeneities of
extinction we compared standard deviations of the LMC $P-L$ relation
constructed for observed magnitudes with those presented below for
extinction corrected samples. We found improvement of the standard
deviation from $\sigma=0.123$ to $\sigma=0.109$ in the {\it I}-band and
$\sigma=0.183$ to $\sigma=0.159$ in the {\it V}-band for the observed
and extinction corrected samples, respectively. This indicates that our
extinction procedure works correctly.  In the SMC the decrease of
standard deviation is smaller due to more uniform extinction there. 

It should be, however, stressed here that extinction is variable within
each Cloud, growing with the distance inside the Cloud. Therefore in any
line-of-sight we may expect an additional scatter of brightness when the
mean extinction correction is used. Also the extinction correction we
applied was determined from different population of stars than Cepheids,
namely old red clump stars. It is possible that the spatial distribution
of red clump stars along the line-of-sight is different than that of
much younger Cepheids. Thus, some systematic differences between the
determined reddening and the mean reddening of Cepheids in a given
direction cannot be ruled out. 

After correcting photometry of our samples of Cepheids for interstellar
extinction we determined the $P-L-C$ and $P-L$ relation with an
iterative procedure using the least square method. We fitted the
relations in the following form:

$$ M=\alpha\cdot \log P + \beta\cdot CI + \gamma \eqno{(1)} $$

\noindent
for the $P-L-C$ relation, and 

$$ M=a\cdot \log P + b \eqno{(2)} $$

\noindent
for the $P-L$ relation.  $P$ is the period of Cepheid, $M$  magnitude
and $CI$ color index.

After each iteration the points deviating by more than $2.5\sigma$ were
removed and the fitting was repeated. In this way we removed a few
outliers -- usually the objects reddened significantly more than the
mean correction applied to our data or objects blended and unresolved
with background stars. $2.5\sigma$ value was selected after a few tests
as a compromise value allowing effective removing of outliers and on the
other hand not removing too many good objects.

The $P-L-C$ relation was determined for the {\it I}-band brightness and
$V-I$ color. The $P-L$ relation was constructed for the {\it B,V} and
{\it I}-bands. Additionally we determined the $P-L$ relation for
extinction insensitive index $W_I$ (called sometimes Wesenheit index,
Madore and Freedman 1991) which is defined as follows:

$$ W_I=I-1.55*(V-I) \eqno{(3)}  $$  

The coefficient 1.55 in Eq.~3 corresponds to the coefficient resulting
from standard interstellar extinction curve dependence of the {\it
I}-band extinction on $E(V-I)$ reddening (Schlegel, Finkbeiner and Davis
1998). It is easy to show that the values of $W_I$  are the same when
derived from observed or extinction free magnitudes, provided that
extinction to the object is not too high so it can be approximated with
a linear function of color.

The {\it B}-band $P-L$ relation is presented for the SMC only. As we
mentioned in Section~2, the LMC sample is less complete in the {\it
B}-band because only 8 from 21 fields from the LMC have already been
reduced in the {\it B}-band. Complete sample will be available after the
next observing season of the LMC. The {\it B}-band sample of the LMC
Cepheids presently at our disposal is less than  half that numerous than
in the {\it VI}-bands. In particular the longer period Cepheids, very
important for precise determination of the $P-L-C$ and $P-L$ relations,
are sparsely populated what would make our determinations less accurate.
Therefore to avoid biases we decided to wait with determination of the
{\it B}-band $P-L$ relation for the LMC until the full data set is
available. One has to remember that the {\it B}-band $P-L$ relation  has
intrinsically much larger scatter what makes it much less attractive for
distance determination (\cf the SMC data).  Also, all most important data
for extragalactic Cepheids collected by the HST Key Project team were
obtained in the bands closely resembling standard {\it VI}-bands, thus
precise calibration of $P-L$ relations in these bands is more important.

In this paper we concentrate on analysis of the $P-L$ and $P-L-C$
diagrams of the more important FU mode Cepheids. Cepheids of this type
are usually discovered in extragalactic objects and used for distance
determination. We limited our samples of FU mode Cepheids from the short
period side at $\log P=0.4$ for two reasons. First,  in the LMC the
population of Cepheids with shorter periods is marginal contrary to the
SMC where large sample of Cepheids with periods shorter than $\log
P=0.4$ (2.5 days) has been found. Therefore to be able to make
non-biased comparison of our relations in both galaxies we limited
ourselves to the same range of periods in both galaxies. Secondly, Bauer
\etal (1999) reported that the slope of the fundamental mode pulsators
with periods shorter than 2 days in the SMC is steeper than for the
longer period stars. Indeed, we also observe in our data such a change
of slope. Our lower limit of $\log P=0.4$ excludes safely this part of
the $P-L$ relation of FU mode Cepheids in the SMC. The upper limit of
period of our samples is defined by saturation level of the CCD detector
because the longest period Cepheids become too bright and are
overexposed in our images. It is at $\log P\approx1.5$ and $\log
P\approx1.7$ for the LMC and SMC, respectively.

\MakeTable{crrrr}{7cm}{Best least square fit parameters of the $P-L$ relation:
$M=a\cdot\log P + b$}
{
\hline
\hline
\noalign{\vskip2pt}
\multicolumn{5}{c}{LMC -- Fundamental Mode Cepheids}\\
\noalign{\vskip2pt}
\hline
\hline
\noalign{\vskip2pt}
Band        &\multicolumn{1}{c}{$a$} &\multicolumn{1}{c}{$b$}& 
\multicolumn{1}{c}{$N$}&\multicolumn{1}{c}{$\sigma$}\\
\noalign{\vskip2pt}
\hline
\noalign{\vskip2pt}
$I_0$       &$-2.962$      & 16.558      & 658         & 0.109\\
            & 0.021        &  0.014      &             &      \\
$V_0$       &$-2.760$      & 17.042      & 649         & 0.159\\
            & 0.031        &  0.021      &             &      \\
$W_I$       &$-3.277$      & 15.815      & 690         & 0.076\\
            & 0.014        &  0.010      &             &      \\
\hline
\hline
\noalign{\vskip2pt}
\multicolumn{5}{c}{SMC -- Fundamental Mode Cepheids}\\
\noalign{\vskip2pt}
\hline
\hline
\noalign{\vskip2pt}
Band        &\multicolumn{1}{c}{$a$} &\multicolumn{1}{c}{$b$}& 
\multicolumn{1}{c}{$N$}&\multicolumn{1}{c}{$\sigma$}\\
\noalign{\vskip2pt}
\hline
\noalign{\vskip2pt}
$I_0$       &$-2.857$      & 17.039      & 488         & 0.205\\
            & 0.033        &  0.025      &             &      \\
$V_0$       &$-2.572$      & 17.480      & 466         & 0.257\\
            & 0.042        &  0.032      &             &      \\
$B_0$       &$-2.207$      & 17.711      & 465         & 0.319\\
            & 0.053        &  0.041      &             &      \\
$W_I$       &$-3.303$      & 16.345      & 469         & 0.135\\
            & 0.022        &  0.017      &             &      \\
\hline
\hline
}

\Section{Discussion}

Tables~1 and 3 present results of the least-square (LSQ) fitting of the
$P-L$ and $P-L-C$ relations to our samples of classical, fundamental
mode  Cepheids,  respectively. We also list there number of stars used
and standard deviation of the residual magnitudes.

We tested stability of our best LSQ solutions performing a few
simulations. First, we limited the samples by cutting the lower period
limit. Then, we removed randomly significant number (up to the half) of
shorter period ($\log P < 0.7$) Cepheids which in both samples are much
more numerous than the longer period ones. In all cases results were
consistent with our best fits for the entire samples differing by not
more than $\pm0.05$ in $\log P$ coefficients of our relations.

We also checked whether our sample of the LMC Cepheids is not severely
affected by differential extinction. We performed a series of tests by
limiting the sample of  Cepheids to those from the fields in which there
are indications that the extinction is in the first approximation
uniform. The shape of the red clump in the color-magnitude diagram of a
given field served as an indicator of how uniform the extinction in the
field is. In  many fields the shape of the red clump is round indicating
little differential extinction. However, in a few cases the oval shape
of the red clump, elongated in the direction of reddening, clearly
indicates larger differential extinction. We excluded Cepheids located
in these fields from our sample. This lowered the number of objects from
about 690 to 480. Fitting the $P-L$ and $P-L-C$ relations to such a
cleaned sample gave almost identical results as for the full sample for
all combination of bands and relations. Thus, our tests indicate that
the differential extinction is of little concern in our case.

Simple comparison of results obtained for the LMC and SMC Cepheids
allows to draw some conclusions on possible dependence of the $P-L$ 
relation on differences of metallicity of these galaxies. We find from
Table~1 that the slopes, $a$, of the $P-L$ relation  are within errors
the same for the {\it I}-band and $W_I$ index for the LMC and SMC.  Only
in the case of the {\it V}-band $P-L$ relation we note marginally 
shallower slope ($3.6\sigma$) for the SMC relation as compared to the
LMC one. However, taking into account uncertainty of the true shape of
Cepheid extinction which might, for instance, depend slightly on the
brightness of Cepheid, the same coefficients for extinction insensitive
index $W_I$ and, finally, larger dispersion of the $P-L$ relations in
the {\it V}-band  we do not consider this somewhat lower slope in the
SMC as significant. We may conclude that for the metallicity range
between the LMC and SMC (${\rm [Fe/H]}=-0.3$~dex, and $-0.7$~dex for the
LMC and SMC, respectively, Luck \etal 1998) the slopes, $a$, of the
$P-L$ relations of fundamental mode classical Cepheids are within errors
constant.

\MakeTable{crrrr}{8cm}{Final, adopted parameters of the $P-L$
relation: $M=a\cdot\log P + b$}
{
\hline
\hline
\noalign{\vskip2pt}
\multicolumn{5}{c}{LMC -- Fundamental Mode Cepheids}\\
\noalign{\vskip2pt}
\hline
\noalign{\vskip2pt}
Band        &\multicolumn{1}{c}{$a$} &\multicolumn{1}{c}{$b$}& 
\multicolumn{1}{c}{$N$}&\multicolumn{1}{c}{$\sigma$}\\
\noalign{\vskip2pt}
\hline
\noalign{\vskip2pt}
$I_0$       &$-2.962$      & 16.558      & 658         & 0.109\\
$V_0$       &$-2.760$      & 17.042      & 649         & 0.159\\
$W_I$       &$-3.277$      & 15.815      & 690         & 0.076\\
\hline
\hline
\noalign{\vskip2pt}
\multicolumn{5}{c}{SMC -- Fundamental Mode Cepheids}\\
\noalign{\vskip2pt}
\hline
\noalign{\vskip2pt}
Band        &\multicolumn{1}{c}{$a$} &\multicolumn{1}{c}{$b$}& 
\multicolumn{1}{c}{$N$}&\multicolumn{1}{c}{$\sigma$}\\
\noalign{\vskip2pt}
\hline
\noalign{\vskip2pt}
$I_0$       &$-2.962$      & 17.114      & 488         & 0.207\\
$V_0$       &$-2.760$      & 17.611      & 466         & 0.258\\
$W_I$       &$-3.277$      & 16.326      & 469         & 0.136\\
\hline
\hline
}

Because the $P-L$ relations of the LMC have much smaller scatter (the
standard deviation is almost two times smaller for the LMC relations as
compared to the SMC ones) and in the case of the fundamental mode
Cepheids of $\log P > 0.4$ they are much better populated, we decided to
use the coefficients $a$ derived from the LMC data as universal and we
repeated fitting of the SMC data with these coefficients. The fits we
obtained are only slightly worse than the best LSQ fits and we treat
them as final. Adopted parameters of the $P-L$ relations for the LMC and
SMC are listed in Table~2.

We should note at this point that fitting of the $P-L$ relation for the
first overtone Cepheids leads to somewhat different results. Although
the {\it V} and {\it I}-band slopes are within errors the same, the 
extinction insensitive index $W_I$ indicates a small
difference of slopes  of its $P-L$ relation in the LMC and SMC at the
$4.6\sigma$ level ($a_{W_I}^{LMC}=-3.406\pm0.021, 
a_{W_I}^{SMC}=-3.556\pm0.025$). Because the first overtone Cepheids are
not used as distance indicators, we only note this small discrepancy,
but the problem certainly deserves further studies.

\MakeTable{crrrrr}{8cm}{Best least square fit parameters of the $P-L-C$ relation:
$I_0=\alpha\cdot\log P+ \beta\cdot(V-I)_0+\gamma$}
{
\hline
\hline
\noalign{\vskip2pt}
\multicolumn{6}{c}{LMC -- Fundamental Mode Cepheids}\\
\noalign{\vskip2pt}
\hline
\hline
\noalign{\vskip2pt}
Band        &\multicolumn{1}{c}{$\alpha$} &\multicolumn{1}{c}{$\beta$}
&\multicolumn{1}{c}{$\gamma$}& \multicolumn{1}{c}{$N$}&\multicolumn{1}{c}{$\sigma$}\\
\noalign{\vskip2pt}
\hline
\noalign{\vskip2pt}
$I_0$         &$-3.246$      & 1.409   & 15.884      & 685         & 0.074\\
              &  0.015       & 0.026   &  0.016      &             &      \\
\hline
\hline
\noalign{\vskip2pt}
\multicolumn{6}{c}{SMC -- Fundamental Mode Cepheids}\\
\noalign{\vskip2pt}
\hline
\hline
\noalign{\vskip2pt}
Band        &\multicolumn{1}{c}{$\alpha$} &\multicolumn{1}{c}{$\beta$}
&\multicolumn{1}{c}{$\gamma$}& \multicolumn{1}{c}{$N$}&\multicolumn{1}{c}{$\sigma$}\\
\noalign{\vskip2pt}
\hline
\noalign{\vskip2pt}
$I_0$         &$-3.487$      & 2.116   & 16.107      & 464         & 0.126\\
              &  0.027       & 0.062   &  0.032      &             &      \\
\hline
\hline
}

Table~3 presents results of the best LSQ fitting of the LMC and SMC
$P-L-C$ relations.  Comparison of these relations in the LMC and SMC
requires special attention. At the first look it may seem that
coefficients $\alpha$ and $\beta$ are different by many sigmas as the
direct comparison of figures in Table~3 indicates. However, it does not
necessarily mean that they are indeed different. It is well known that
the $\alpha$ and $\beta$ coefficients of the $P-L-C$ relation are highly
correlated in the sense that the error in $\alpha$ coefficient is
coupled with the error of the $\beta$ coefficient and both errors
compensate (Caldwell and Coulson 1986). This makes precise empirical
determination of both coefficients difficult but has little consequences
on predicted luminosity (and distance determination) if both
coefficients come from the same determination. Thus, the $\alpha$ and
$\beta$ should be considered as a pair. To investigate whether the SMC
data can be approximated well by the LMC pair of coefficients ($\alpha$,
$\beta$), we repeated the $P-L-C$ fitting of the SMC data with $\alpha$
and $\beta$ fixed from the LMC determination. Results are given in
Table~4 and Fig.~1 which presents the $P-L-C$ relation in the form of
plot of $I_0-1.409\cdot(V-I)_0$ against $\log P$ for both the LMC and
SMC. The fit is somewhat worse than the best LSQ fit of the SMC data
($\sigma=0.126$ \vs 0.138, for the best LSQ and LMC coefficients fit,
respectively), but it is clearly seen from Fig.~1, that the pair
($\alpha$,$\beta$) from the LMC fits the SMC data almost equally well.
Both sequences of Cepheids for the LMC and SMC in Fig.~1 are within
errors  parallel indicating that differences between the best LSQ fit
and that with LMC coefficients ($\alpha$, $\beta$) are marginal.
Therefore, there is no indication that these coefficients significantly
differ between the LMC and SMC.

\MakeTable{crrrrr}{8cm}{Final, adopted parameters of the $P-L-C$
relation: $I_0=\alpha\cdot\log P+ \beta\cdot(V-I)_0+\gamma$}
{
\hline
\hline
\noalign{\vskip2pt}
\multicolumn{6}{c}{LMC -- Fundamental Mode Cepheids}\\
\noalign{\vskip2pt}
\hline
\noalign{\vskip2pt}
Band        &\multicolumn{1}{c}{$\alpha$} &\multicolumn{1}{c}{$\beta$}
&\multicolumn{1}{c}{$\gamma$}& \multicolumn{1}{c}{$N$}&\multicolumn{1}{c}{$\sigma$}\\
\noalign{\vskip2pt}
\hline
\noalign{\vskip2pt}
$I_0$       &$-3.246$      & 1.409   & 15.884     & 685        & 0.074\\
\hline
\hline
\noalign{\vskip2pt}
\multicolumn{6}{c}{SMC -- Fundamental Mode Cepheids}\\
\noalign{\vskip2pt}
\hline
\noalign{\vskip2pt}
Band        &\multicolumn{1}{c}{$\alpha$} &\multicolumn{1}{c}{$\beta$}
&\multicolumn{1}{c}{$\gamma$}& \multicolumn{1}{c}{$N$}&\multicolumn{1}{c}{$\sigma$}\\
\noalign{\vskip2pt}
\hline
\noalign{\vskip2pt}
$I_0$       &$-3.246$      & 1.409   & 16.399 & 463        & 0.138\\
\hline
\hline
}

\begin{figure}[htb]
\hglue-0.5cm\psfig{figure=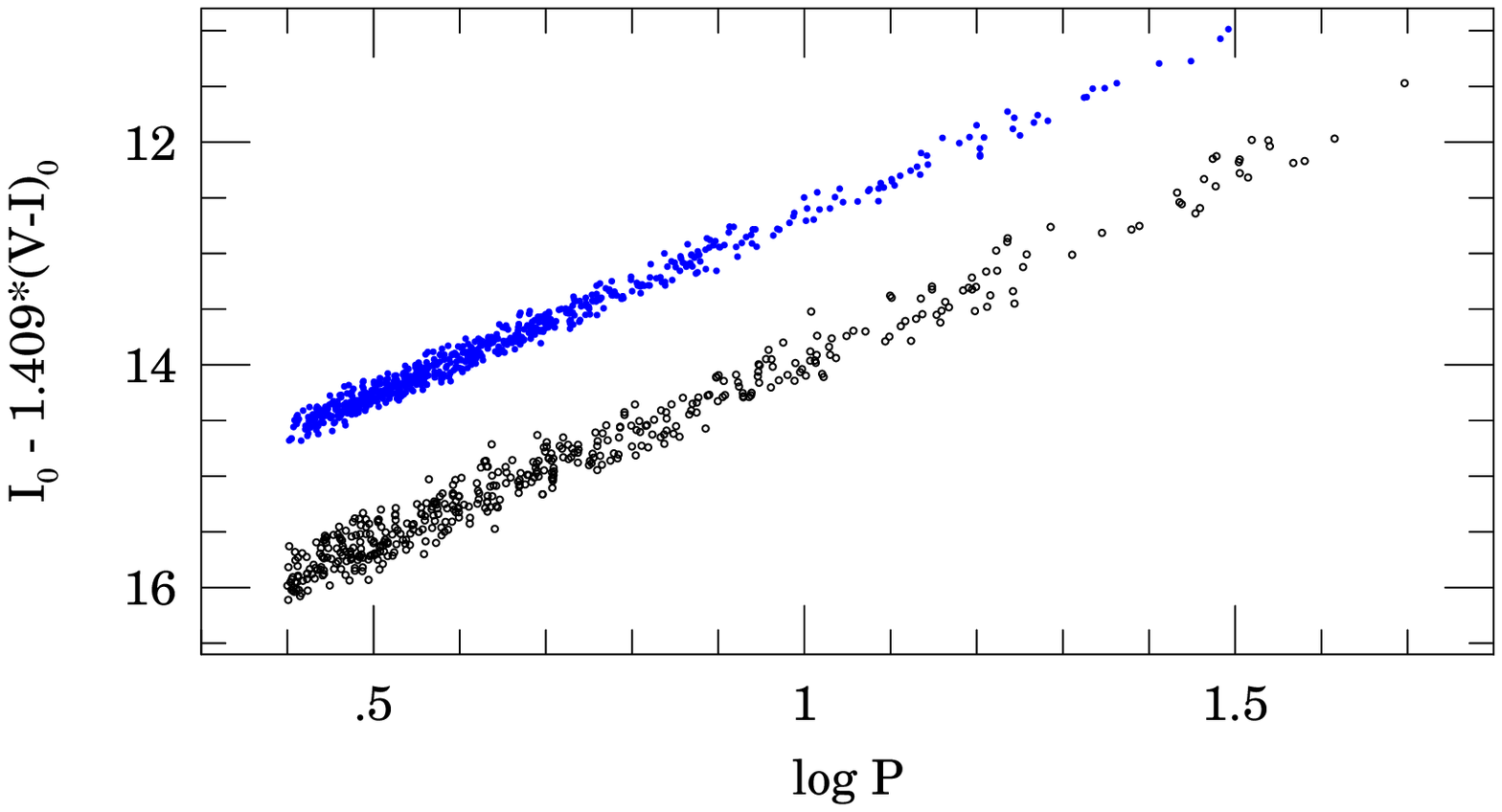,bbllx=30pt,bblly=140pt,bburx=505pt,bbury=410pt,width=13cm,clip=}
\vspace*{3pt}
\FigCap{{\it I}-band magnitudes of the fundamental mode Cepheids with
subtracted color term resulting from the final $P-L-C$ relation (Table~4).
Filled dots and open circles correspond to the LMC and SMC Cepheids,
respectively. The SMC diagram is shifted additionally down by 0.8~mag.}
\end{figure}

Final, adopted parameters of the $P-L-C$ relation in the LMC and SMC are
provided in Table~4. It is worth noting that the color term, $\beta$, of
the LMC determination (and as a consequence the SMC, because we adopt
the LMC values of $\alpha$ and $\beta$ as universal) is very close to
the coefficient of the {\it I}-band extinction dependence on $E(V-I)$,
1.55, making the fitting of the $P-L-C$ relation non-sensitive to
interstellar extinction. 

Figures 2--4 and 5--8 show the $P-L$ relations for LMC and SMC Cepheids,
respectively, for the {\it BVI}-bands and $W_I$ index.  In the upper
panel of each figure all observed Cepheids are plotted. Dark and light
dots indicate the fundamental mode and first overtone Cepheids,
respectively. In the lower panel, the $P-L$ relation for the fundamental
mode Cepheids is shown.  Dark and light points in the lower panel
mark objects included and rejected from the final fits,
respectively. Solid line shows the $P-L$ relation with coefficients
adopted from Table~2.

The $P-L-C$ relation (Fig.~1) and $W_I$ index $P-L$ relation (Fig.~4) of
the LMC show incredibly small scatter from the fitted relation. The
standard deviation of differences between the observed and fitted values
is equal to only 0.074~mag and it determines most likely the intrinsic
dispersion of the $P-L-C$ and $P-L$ relations. It also proves that the
Cepheid variable stars can indeed be  a very good standard candle
allowing precise distance determination good to a few percent. In the
case of the SMC the scatter is somewhat larger amounting to
$\sigma=0.126$~mag. This could be expected as it is widely believed that
the geometrical depth of the SMC, which is tilted to the line of sight
much more than seen almost face-on LMC, is larger than that of  the LMC
(Caldwell and Coulson 1986).

\begin{figure}[htb]
\hglue-0.5cm\psfig{figure=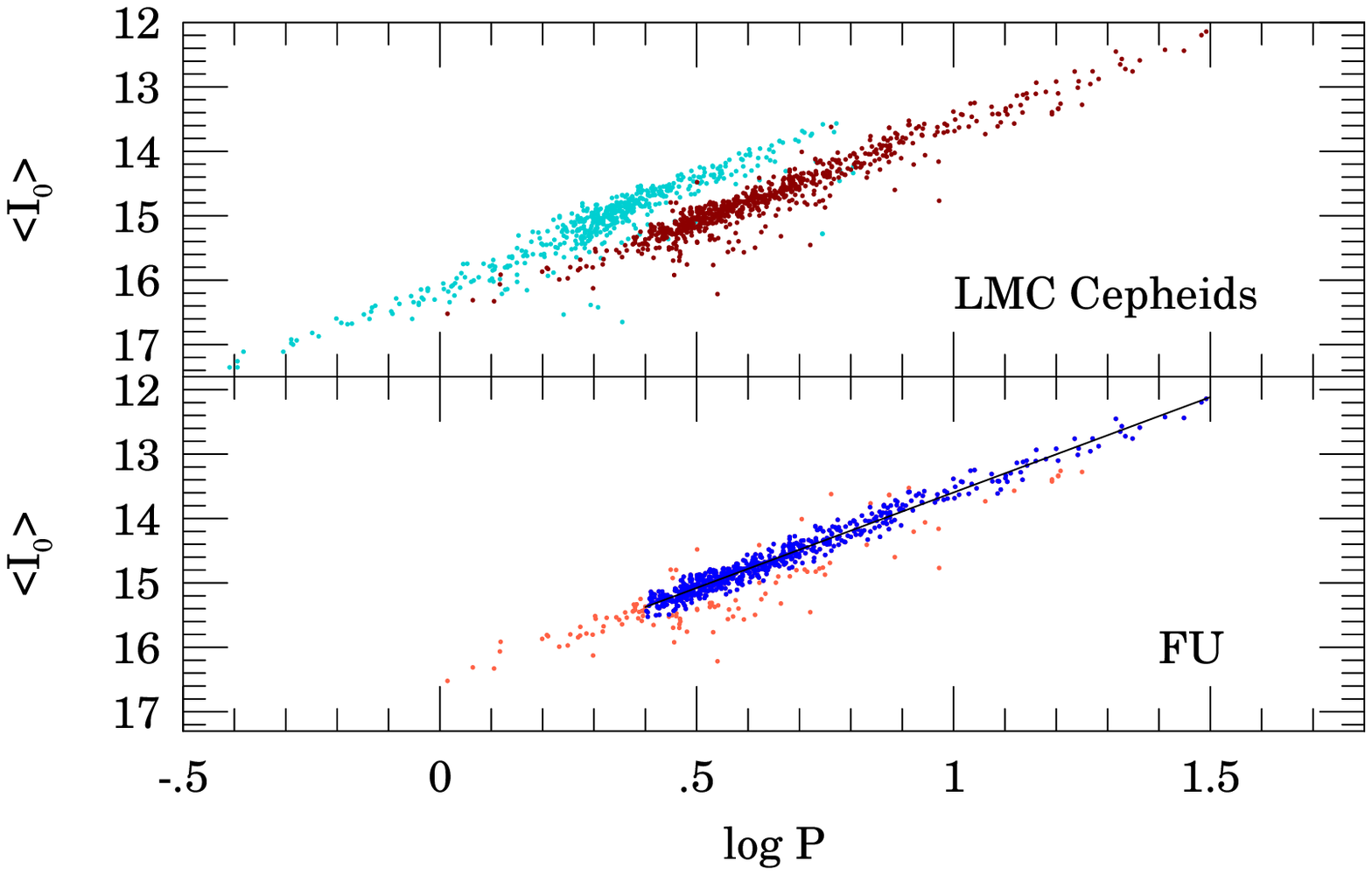,bbllx=30pt,bblly=225pt,bburx=505pt,bbury=530pt,width=13cm,clip=}
\vspace*{3pt}
\FigCap{Upper panel: {\it I}-band $P-L$ relation of the LMC Cepheids.
Darker and lighter dots indicate FU and FO mode Cepheids, respectively.
Lower panel: $P-L$ relation for the FU Cepheids. Solid line indicates
adopted approximation (Table~2). Dark and light dots correspond to stars
used and rejected in the final fit, respectively. }
\end{figure}

\begin{figure}[p]
\hglue-0.5cm\psfig{figure=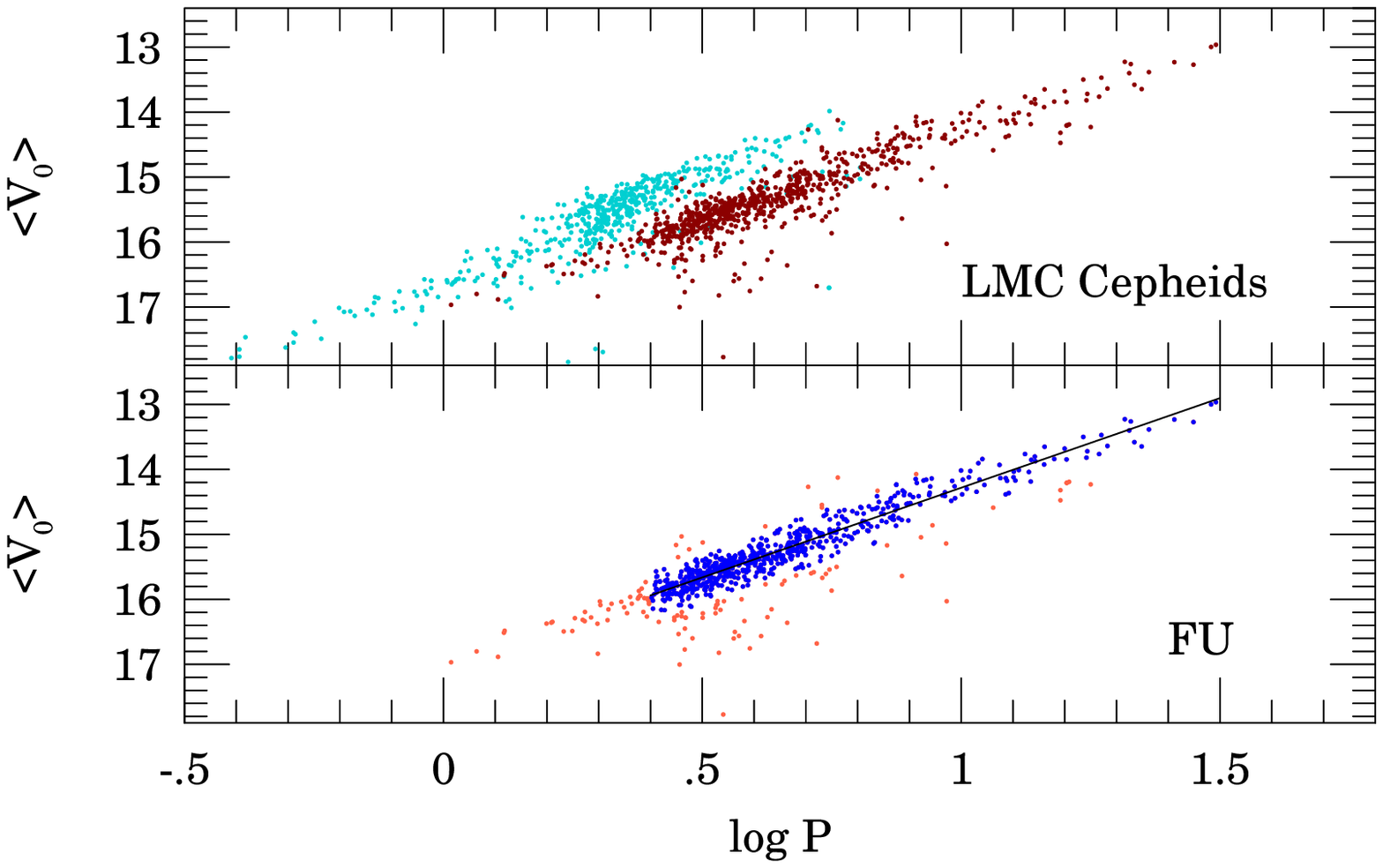,bbllx=30pt,bblly=225pt,bburx=505pt,bbury=530pt,width=13cm,clip=}
\FigCap{Upper panel: {\it V}-band $P-L$ relation of the LMC Cepheids.
Darker and lighter dots indicate FU and FO mode Cepheids, respectively.
Lower panel: $P-L$ relation for the FU Cepheids. Solid line indicates
adopted approximation (Table~2). Dark and light dots correspond to stars
used and rejected in the final fit, respectively. }
\vspace*{3pt}
\hglue-0.5cm\psfig{figure=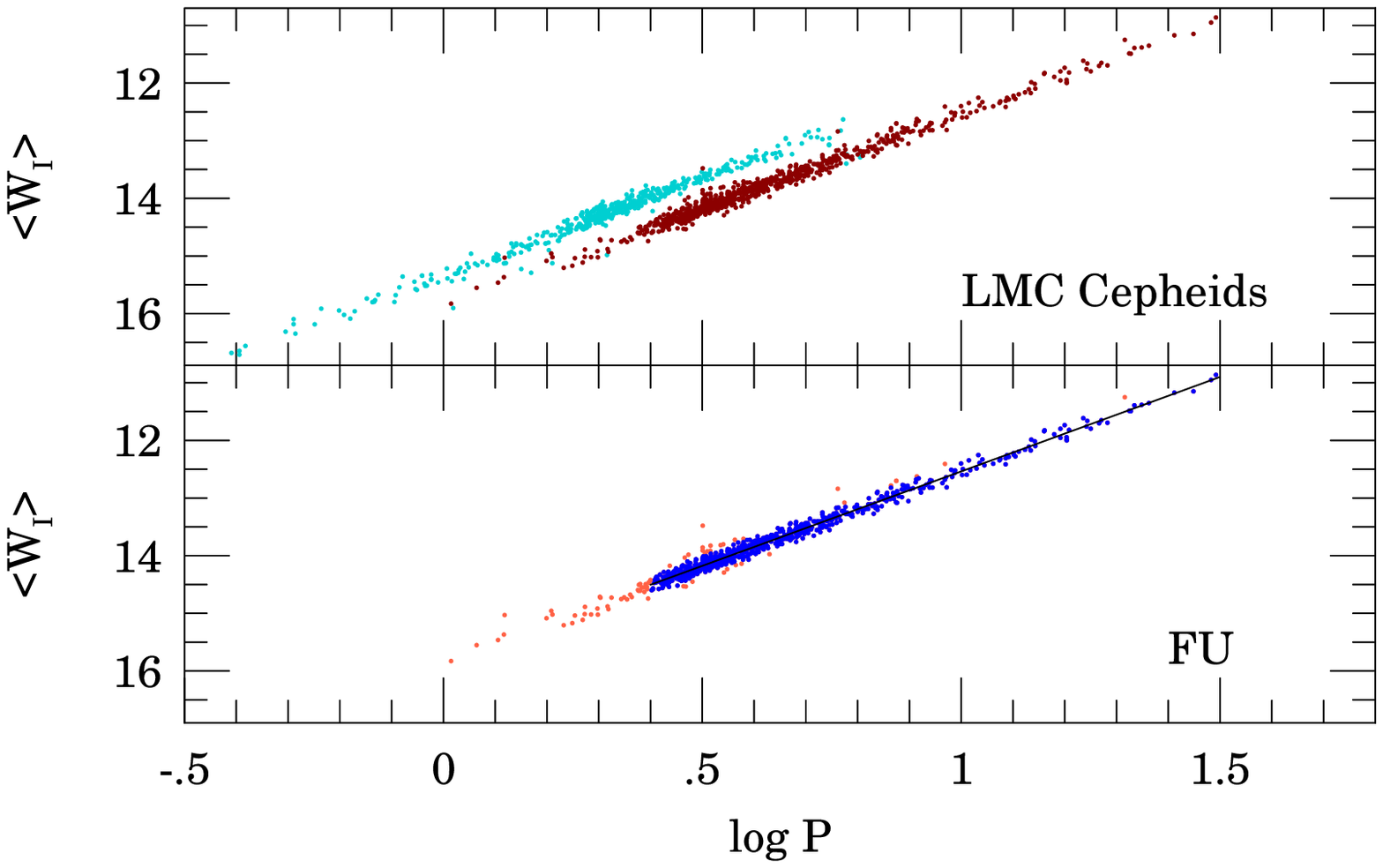,bbllx=30pt,bblly=225pt,bburx=505pt,bbury=530pt,width=13cm,clip=}
\vspace*{3pt}
\FigCap{Upper panel: $W_I$ index $P-L$ relation of the LMC Cepheids.
Darker and lighter dots indicate FU and FO mode Cepheids, respectively.
Lower panel: $P-L$ relation for the FU Cepheids. Solid line indicates
adopted approximation (Table~2). Dark and light dots correspond to stars
used and rejected in the final fit, respectively. }
\end{figure}

\begin{figure}[p]
\hglue-0.5cm\psfig{figure=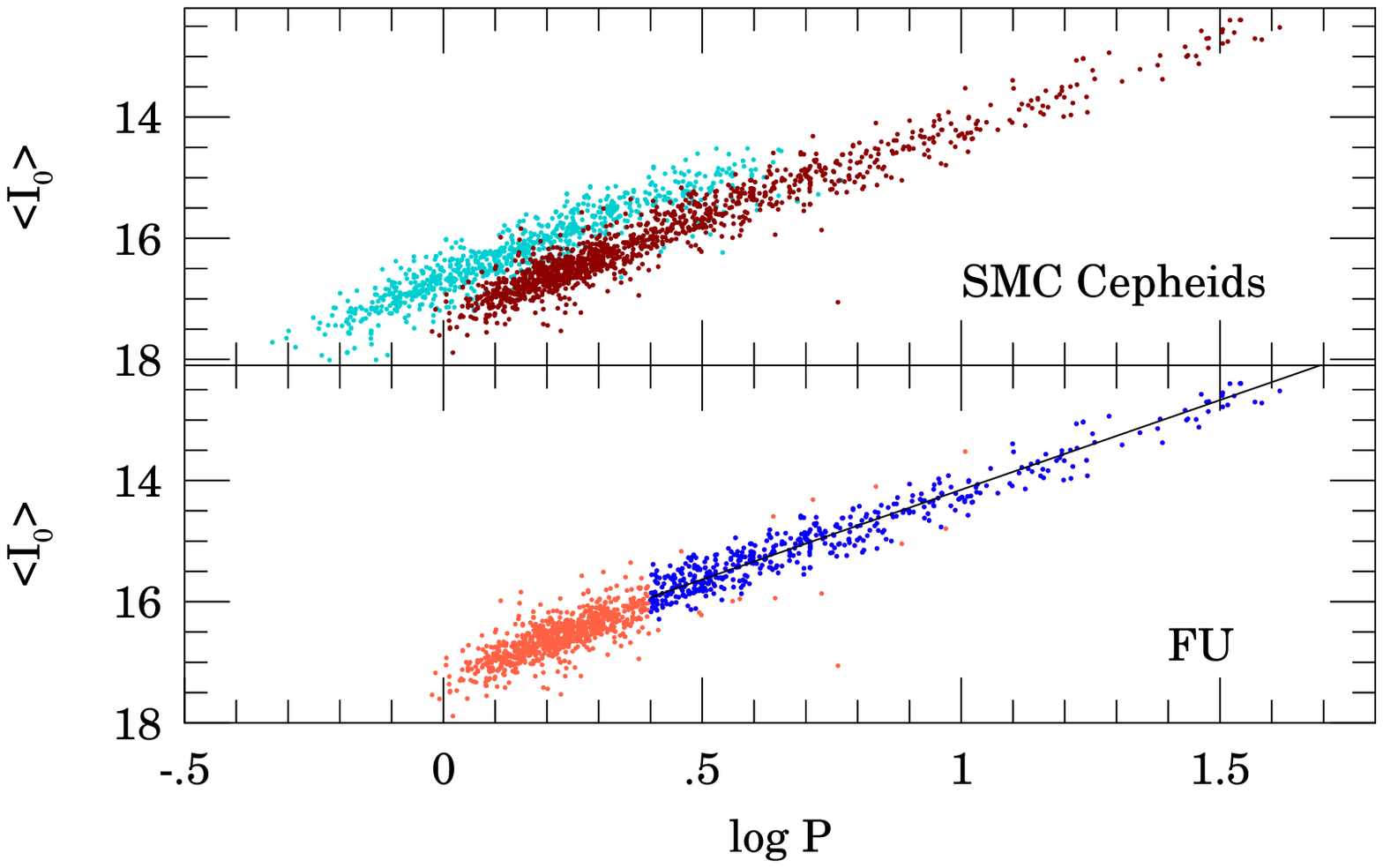,bbllx=30pt,bblly=225pt,bburx=505pt,bbury=530pt,width=13cm,clip=}
\FigCap{Upper panel: {\it I}-band $P-L$ relation of the SMC Cepheids.
Darker and lighter dots indicate FU and FO mode Cepheids, respectively.
Lower panel: $P-L$ relation for the FU Cepheids. Solid line indicates
adopted approximation (Table~2). Dark and light dots correspond to stars
used and rejected in the final fit, respectively. }
\vspace*{3pt}
\hglue-0.5cm\psfig{figure=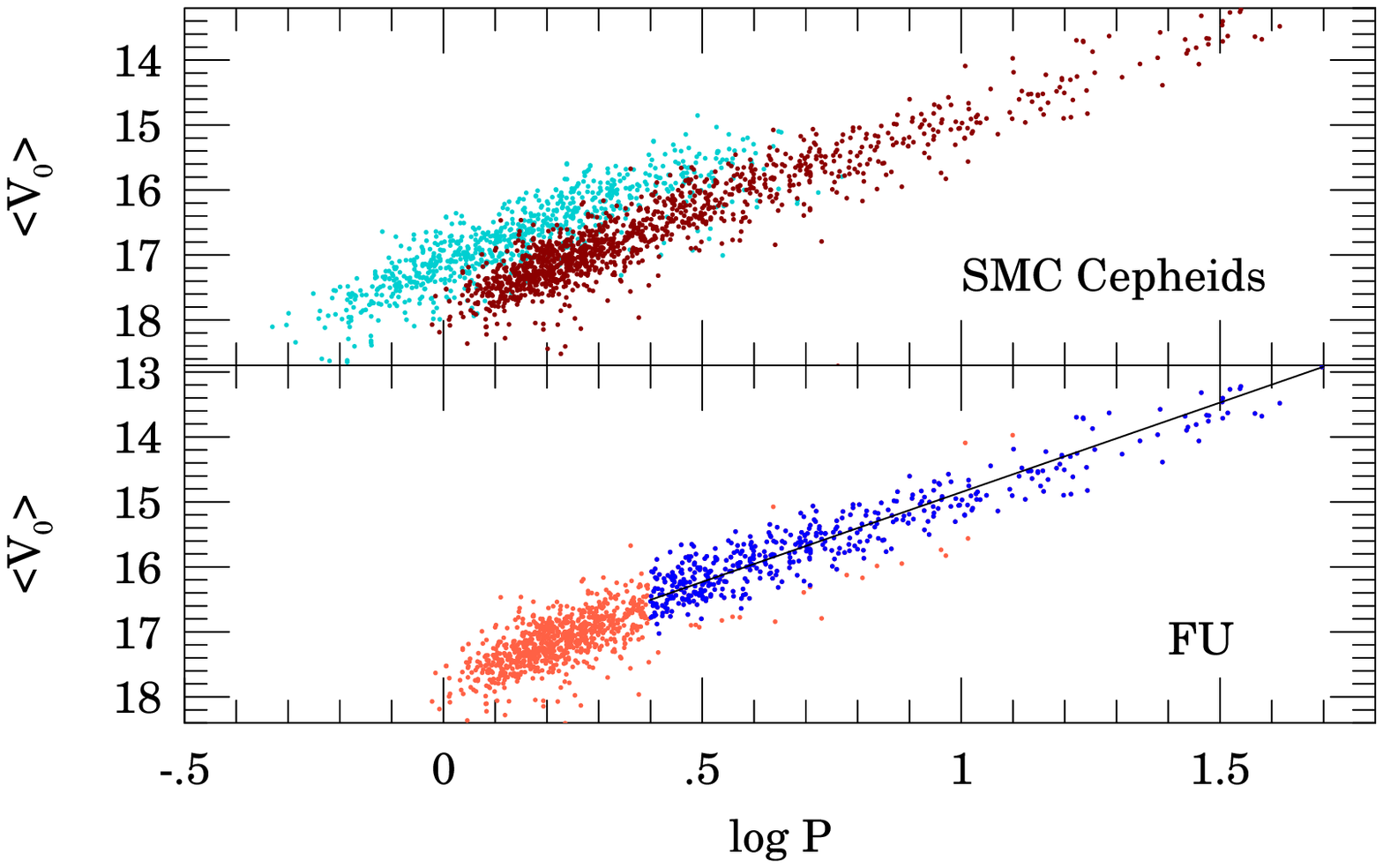,bbllx=30pt,bblly=225pt,bburx=505pt,bbury=530pt,width=13cm,clip=}
\FigCap{Upper panel: {\it V}-band $P-L$ relation of the SMC Cepheids.
Darker and lighter dots indicate FU and FO mode Cepheids, respectively.
Lower panel: $P-L$ relation for the FU Cepheids. Solid line indicates
adopted approximation (Table~2). Dark and light dots correspond to stars
used and rejected in the final fit, respectively. }
\end{figure}

\begin{figure}[p]
\hglue-0.5cm\psfig{figure=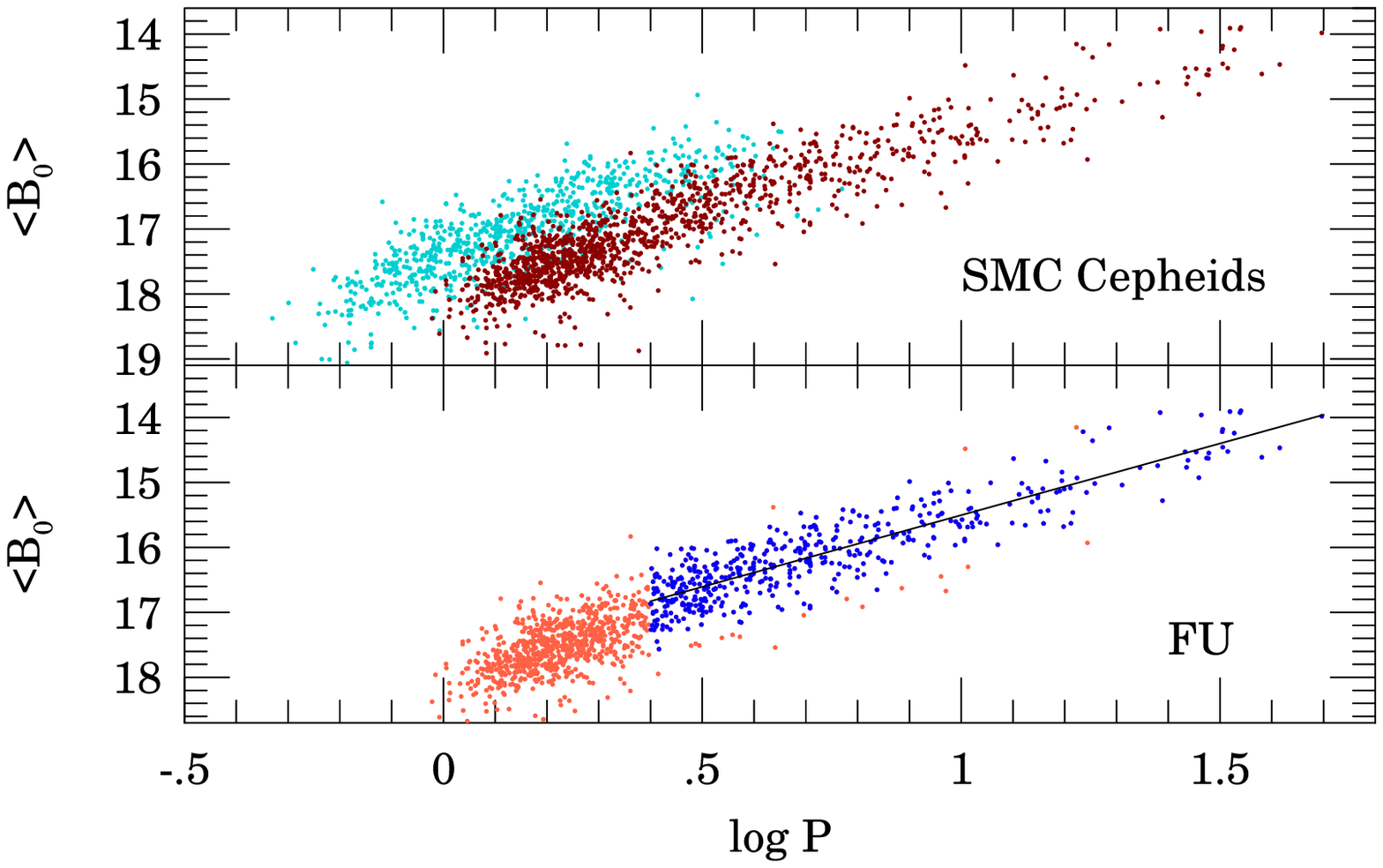,bbllx=30pt,bblly=225pt,bburx=505pt,bbury=530pt,width=13cm,clip=}
\FigCap{Upper panel: {\it B}-band $P-L$ relation of the SMC Cepheids.
Darker and lighter dots indicate FU and FO mode Cepheids, respectively.
Lower panel: $P-L$ relation for the FU Cepheids. Solid line indicates
adopted approximation (Table~1). Dark and light dots correspond to stars
used and rejected in the final fit, respectively. }
\vspace*{3pt}
\hglue-0.5cm\psfig{figure=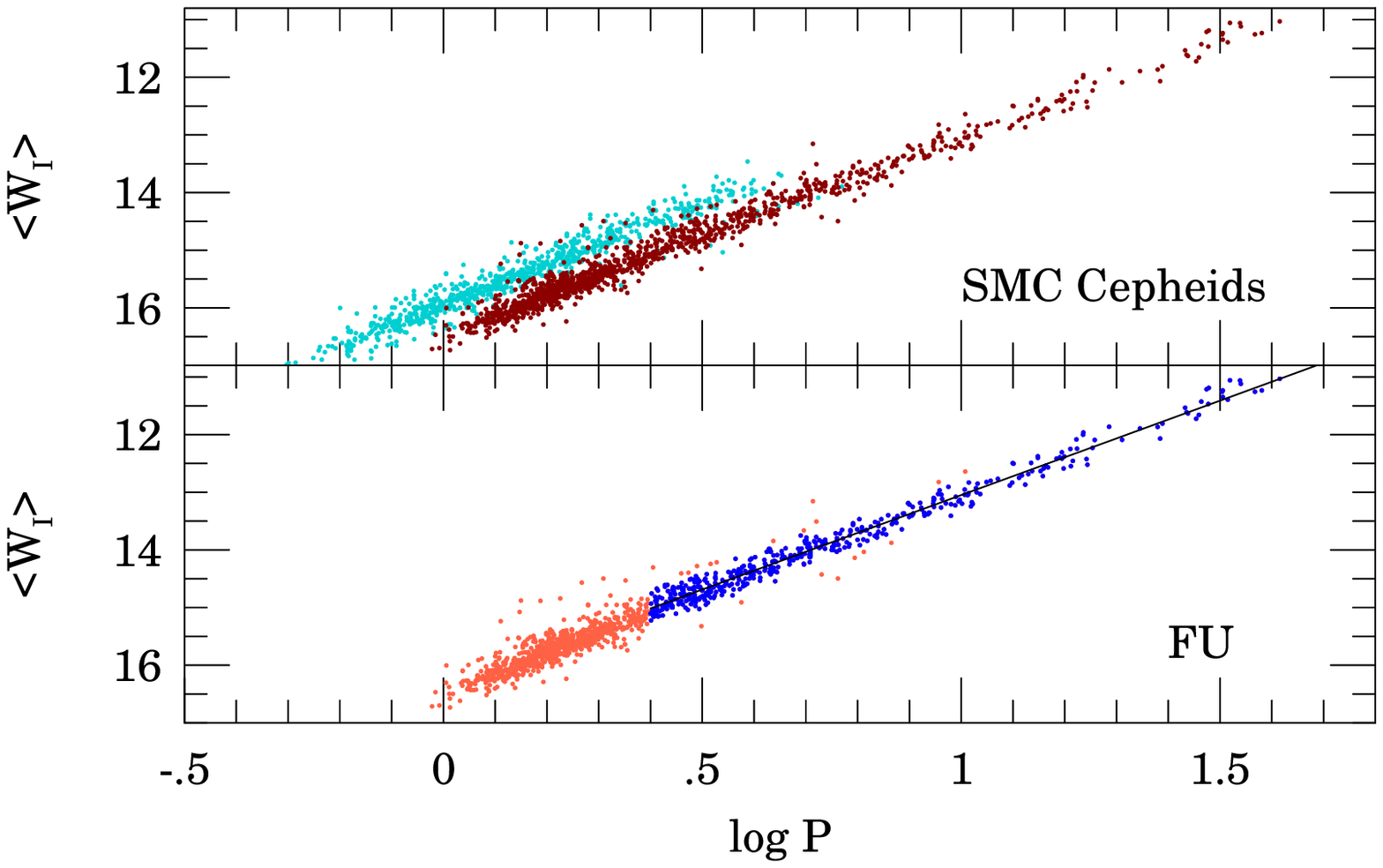,bbllx=30pt,bblly=225pt,bburx=505pt,bbury=530pt,width=13cm,clip=}
\vspace*{3pt}
\FigCap{Upper panel: $W_I$ index $P-L$ relation of the SMC Cepheids.
Darker and lighter dots indicate FU and FO mode Cepheids, respectively.
Lower panel: $P-L$ relation for the FU Cepheids. Solid line indicates
adopted approximation (Table~2). Dark and light dots correspond to stars
used and rejected in the final fit, respectively. }
\end{figure}

\Subsection {SMC -- LMC Distance Ratio}

Comparisons of coefficients $a$, of the $P-L$ relations and ($\alpha$,
$\beta$) of the $P-L-C$ relation in the LMC and SMC indicate no
significant difference of their values in these galaxies. This is in
agreement with most of  theoretical modeling (Chiosi, Wood and Capitanio
1993, Saio and Gautschy 1998,  Alibert \etal 1999) although the opposite
predictions can also be found in the literature (Bono \etal 1999).

On the other hand, it is believed that metallicity variations among
objects may have stronger effect on the zero point of $P-L-C$ or
$P-L$ relations. Results of previous empirical attempts to determine
effects of metallicity on zero points, generally suggested fainter
Cepheids in metal poorer objects, however,  with high degree of
uncertainty (Sasselov \etal 1997, Kochanek 1997, Kennicutt \etal 1998). 

We may test the dependence of the zero points of $P-L-C$ or $P-L$ relations
on metallicity in very straightforward manner -- by determination of the
difference of  distance moduli between the Magellanic Clouds resulting
from these relations and comparison with similar determinations based on
other reliable distance indicators observed in both Clouds.

With the $P-L$ relation we may determine the difference of distance
moduli of the LMC and SMC for the {\it V} and {\it I}-bands and the
extinction insensitive $W_I$ index. Results -- the difference of zero
points of corresponding $P-L$ relations in the LMC and SMC (Table~2) are
listed in Table~5. We assign lower weight to {\it VI}-band
determinations because of extinction uncertainty.

The distance determined from the $P-L-C$ relation is also extinction
independent. The main source of uncertainty is in this case the color
error. We determined the difference of distance moduli for the period
corresponding to approximately  middle of the relation range: $\log
P=1.0$ for the FU Cepheids. The mean $(V-I)_0$ colors of the  $\log
P=1.0$ fundamental mode Cepheids are $(V-I)_0=0.69\pm0.03$ for the LMC
object and  $(V-I)_0=0.70\pm0.03$ for the SMC Cepheid. Results of
determination with the $P-L-C$ relation are given in Table~5. 

\MakeTable{lccc}{12.5cm}{Difference of distance moduli between the SMC and LMC}
{
\hline
\hline
\noalign{\vskip2pt}
\multicolumn{1}{c}{Method}&   Band   &   $\mu_{SMC}-\mu_{LMC}$& Weight\\
\noalign{\vskip2pt}
\hline
\hline
\noalign{\vskip2pt}
Cepheid $P-L-C$ (FU) &  $I$    & $0.53\pm0.04$ & 1.0\\
Cepheid $P-L$ (FU)   &  $W_I$  & $0.51\pm0.03$ & 1.0\\
Cepheid $P-L$ (FU)   &  $I$    & $0.56\pm0.06$ & 0.5\\
Cepheid $P-L$ (FU)   &  $V$    & $0.57\pm0.08$ & 0.5\\
\noalign{\vskip2pt}
\hline
\noalign{\vskip2pt}
RR~Lyr stars         &  $V$    & $0.51\pm0.08$ & 1.0\\
\noalign{\vskip2pt}
\hline
\noalign{\vskip2pt}
Red Clump stars      &  $I$    & $0.47\pm0.09$ & 1.0\\
\hline
\hline}

To compare results obtained with Cepheids we need independent estimates
of difference of the SMC and LMC distance moduli, derived with other
reliable standard candles observed in both galaxies. We use values
obtained with red clump and RR~Lyr stars. Both types of objects were
observed during the OGLE project with the same equipment and methods of
reductions. In this way possible  systematic errors can be minimized.

Brightness of red clump stars in the LMC and SMC was derived based on
observations of a few star clusters located in the halo of both galaxies
where interstellar extinction is small and can be determined from
reliable maps of Schlegel \etal (1998). The {\it I}-band, mean, 
extinction free  brightness of red clump stars is equal to
$I_0=17.88\pm0.05$ and  $I_0=18.31\pm0.07$ for the LMC and SMC clusters,
respectively (Udalski 1998b). With a small correction for the difference
of metallicity of clusters (Udalski 1998a), the difference of distance
moduli between the LMC and SMC from red clump stars is
$\Delta\mu_{RC}=0.47\pm0.09$~mag. 

Preliminary analysis of RR~Lyr stars from the OGLE fields was presented
in Udalski (1998a). The samples of more than 100 objects in each galaxy
are small as compared to the total number of a few thousands found in
the entire observed area of the Magellanic Clouds, nevertheless they
allow to determine reliable brightness of these objects.  Results for the
LMC RR~Lyr stars presented by Udalski (1998a) were based on moderate
number of observations of these stars and preliminary photometric
calibrations of the observed fields. Also, for both the LMC and SMC RR
Lyr stars, the extinction was estimated by extrapolation or
interpolation of available extinction maps. Now, with about three times
that many observing epochs available for these LMC RR~Lyr stars (about
140 in the {\it I}-band and 20 in the {\it V}-band) and extinction
independently determined we have reanalyzed the objects of Udalski
(1998a).

New photometry of 104 RR~Lyr stars from the LMC with appropriate extinction
correction yields $\langle V^{LMC}_{RR}\rangle =18.94\pm0.04$~mag. This
is practically the same result as presented by Udalski (1998a) ($\langle
V^{LMC}_{RR}\rangle =18.86$~mag), taking into account that extinction
was slightly overestimated in that paper. It is in very good agreement
with the mean brightness of RR~Lyr stars in a few star clusters in
the LMC (Walker 1992). Correction of the mean brightness of RR~Lyr stars
in the SMC is almost negligible: $\langle V^{SMC}_{RR}\rangle
=19.43\pm0.03$~mag as compared to $\langle V^{SMC}_{RR}\rangle
=19.41$~mag in Udalski (1998a).

To derive the difference of distance moduli between both galaxies we
also have to correct the brightness of RR~Lyr stars for metallicity
differences. Fortunately, the mean difference of metallicity of RR~Lyr
stars in the LMC and SMC is not large (on average: ${\rm [Fe/H]}= -1.6 $
and $-1.7$ for the LMC and SMC, respectively, see discussion in Udalski
1998a) and with the average slope of the brightness-metallicity relation
equal to 0.2 it leads to a small correction of 0.02~mag. The SMC RR~Lyr
stars would be fainter if they were of the LMC metallicity. Thus, the
difference of distance moduli between the SMC and LMC resulting from
RR~Lyr stars is equal to $\Delta\mu_{RR}=0.51\pm0.08$~mag. It should be
noted that the mean brightness of RR Lyr stars in both galaxies will be
finally refined when the OGLE catalog of these stars is released.

Results of determinations  of distance moduli difference are summarized
in Table~5. It can be seen that all determinations based on Cepheids are
consistent.  The Cepheid determination is in excellent agreement with
independent estimates from  red clump and RR~Lyr stars. This result
indicates that within uncertainty of a few hundredth of magnitude the
zero points of the $P-L-C$ and $P-L$ relations are independent of
metallicity of hosting object. Thus, the population effects on the
Cepheid distance scale are negligible, at least for the metallicity
range bracketed by the Magellanic Clouds. The mean difference of the SMC
and LMC distance moduli resulting from three independent standard
candles is $\mu_{SMC} - \mu_{LMC}= 0.51\pm0.03$~mag. 

\Subsection{Calibration of the $P-L-C$ and $P-L$ Relations}

The traditional way of calibrating the standard candles is based on
observations of the same type objects located in the Galaxy. In the case
of Galactic Cepheids such a calibration is, however, very difficult.
Even the closest Cepheids are  located, unfortunately, so far from the
Sun that the distance determination can only be performed with indirect
methods. Even Hipparcos did not provide parallaxes precise enough to
allow unambiguous distance determination for sound sample of these
stars. It seems that much better results might be achieved by
observations of Cepheids in other galaxies and calibrating the $P-L$
relation based on other, reliable distance determinations. One of such
galaxies might be NGC4258 to which very precise distance was recently
determined with geometric method, based on maser observations
(Herrnstein \etal 1999). The galaxy possesses a population of Cepheids
detected with HST (Maoz \etal 1999). However, although NGC4258 might be
a very attractive object for testing and checking the calibration of
Cepheids it is certainly not the best object for deriving precise
calibration. The Cepheid sample there is small and it may be biased by
many factors including difficulty of detecting short period Cepheids,
quality of HST photometry  etc.

On the other hand the Magellanic Clouds seem to be the best objects to
calibrate the Cepheid $P-L$ relation. They are close enough and contain
thousands of Cepheids allowing analysis of a large, homogeneous and
photometrically accurate sample of these objects, like the one presented
in this paper. They are also chemically homogeneous what minimizes
uncertainties resulting from metallicity variations (Luck \etal 1998). 

Unfortunately, the distance to the LMC has been a subject of dispute for
a long time. It seems, however,  that the recent results obtained with
different techniques converge at the short distance modulus of
$\mu_{LMC}= 18.2 - 18.3$~mag. The most promising, largely geometric
method using eclipsing binary stars should allow to derive the distance
to the LMC with accuracy of $1-2$ percent. First determination of the
distance to the LMC with HV2274 eclipsing system yields the distance
modulus of $\mu_{LMC}\approx18.26$~mag (Guinan \etal 1998, Udalski \etal
1998c). RR~Lyr stars calibrated with the most reliable methods (Popowski
and Gould 1998) give the distance modulus of
$\mu_{LMC}=18.23\pm0.08$~mag when these calibrations
($M_V^{RR}=0.71\pm0.07$~mag) are used with the mean extinction free {\it
V}-band magnitude of RR~Lyr stars in the LMC presented in the previous
Subsection. Finally, the recent distance determination with  red clump
stars used as a standard candle (Udalski \etal 1998a, Stanek, Zaritsky
and Harris 1998) corrected for small population effects (Udalski
1998a,b) yields the distance modulus of $\mu_{LMC}=18.18\pm0.06$~mag. It
should be noted that  red clump stars are at present the most precisely
calibrated standard candle because  very accurate parallaxes (accuracy
better than 10\%)  were measured for hundreds of them in the solar
neighborhood by Hipparcos.

We should also mention here the  method  that is in principle a precise
geometric technique -- observations of the light echo from the ring of
gas observed around the supernova SN1987A.  Unfortunately in the case of
SN1987A this technique suffers from insufficient quality of observations
at crucial moments after the supernova explosion and many modeling
assumptions. Only the upper limit of the distance modulus to the LMC can
be estimated. It ranges from $\mu_{LMC}<18.58$~mag (Panagia 1998) to as
low as $\mu_{LMC}<18.37$~mag (Gould and Uza 1998).

\MakeTable{crrr}{8cm}{Absolute calibration of the $P-L-C$ and $P-L$
relations with $\mu_{LMC}=18.22$~mag}
{
\hline
\hline
&&&\\
\multicolumn{4}{c}{$P-L-C$ Relation}\\
&&&\\
\hline
\hline
\noalign{\vskip2pt}
\multicolumn{4}{c}{Fundamental Mode Cepheids}\\
\noalign{\vskip2pt}
\hline
\noalign{\vskip2pt}

Band        &\multicolumn{1}{c}{$\alpha$} &\multicolumn{1}{c}{$\beta$}
&\multicolumn{1}{c}{$\gamma$}\\
\noalign{\vskip2pt}
\hline
\noalign{\vskip2pt}
$M_I$         &$-3.246$      & 1.409   & $-2.34$\\
\hline
\hline
&&&\\
\multicolumn{4}{c}{$P-L$ Relation}\\
&&&\\
\hline
\hline
\noalign{\vskip2pt}
\multicolumn{4}{c}{Fundamental Mode Cepheids}\\
\noalign{\vskip2pt}
\hline
\noalign{\vskip2pt}
Band        &\multicolumn{1}{c}{$a$} &\multicolumn{1}{c}{$b$}&\\
\noalign{\vskip2pt}
\hline
\noalign{\vskip2pt}
$M_I$           &$-2.962$      & $-1.66$     &\\
$M_V$           &$-2.760$      & $-1.18$     &\\
$W_{M_I}$       &$-3.277$      & $-2.41$     &\\
\hline
\hline
}

To calibrate the Cepheid $P-L-C$ and $P-L$ relations we adopted the
average distance modulus resulting from all these determinations:
$\mu_{LMC}=18.22\pm0.05$~mag. It is in excellent agreement with the
recent observations of Cepheids in NGC4258 galaxy (Maoz \etal 1999).
Calibrated with the HST Key Project team method  (assuming the distance
modulus to the LMC, $\mu_{LMC}=18.50$) they yield the distance to NGC4258
equal to $8.1\pm0.4$~Mpc, considerably larger ($\approx 12$\%) than
resulting from precise geometric method measurement ($7.2\pm0.3$~Mpc,
Herrnstein \etal 1999). The most natural explanation of this discrepancy
is shorter than assumed distance modulus to the LMC ($\mu_{LMC}\approx
18.2$~mag) -- fully consistent with our adopted value.

If necessary, any refinement of the distance modulus to the LMC  in the
future will correspond to appropriate shift of the zero points of
our calibration. Coefficients of the absolute magnitude $P-L-C$ and
$P-L$ relations for classical, FU mode Cepheids are given in Table~6.

Finally, based on our photometry of Cepheids and RR~Lyr stars in both
Magellanic Clouds we may provide a constraint on the absolute magnitude
of fundamental mode Cepheids. The mean {\it V}-band magnitude of the
10-day period Cepheid  is equal to $\langle V^{LMC}_{C,10}\rangle
=14.28\pm0.03$~mag in the LMC and $\langle V^{SMC}_{C,10}\rangle
=14.85\pm0.04$~mag in the SMC. The extinction free brightness of other
standard candle -- RR~Lyr stars can be used for comparison. In the
previous Subsection we provided appropriate brightness of RR~Lyr stars
in both galaxies: $\langle V^{LMC}_{RR}\rangle =18.94\pm0.04$~mag and 
$\langle V^{SMC}_{RR}\rangle =19.43\pm0.03$~mag for the LMC and SMC,
respectively. Including a small correction of RR~Lyr brightness due to
metallicity differences, we find $\Delta V^{LMC}_{RR-C,10}=4.66\pm0.05$~mag
and $\Delta V^{SMC}_{RR-C,10}=4.60\pm0.05$~mag for the LMC and SMC,
respectively. The difference is in respect to the RR~Lyr star of the LMC
metallicity (${\rm [Fe/H]}=-1.6$~dex). 

Consistent results in both Magellanic Clouds indicate that both Cepheids
and RR~Lyr are good standard candles. Assuming the absolute calibration
of RR~Lyr stars: $M_V^{RR}=0.71\pm0.07$~mag (Popowski and Gould 1998) we
obtain $M_V^{C,10}=-3.92\pm0.09$ for 10-day period Cepheid. Such an
absolute  brightness of Cepheids and our observed $P-L$ relation
($\langle V^{LMC}_{C,10}\rangle =14.28$~mag) indicate the distance
modulus to the LMC $\mu_{LMC}=18.20$~mag fully consistent with the short
distance modulus adopted for absolute calibration {\it via} the LMC
(Table~6). Thus, the distance scale of Cepheids is consistent with
distance scales inferred from RR~Lyr stars and other reliable distance
indicators.

We may also compare the Cepheid absolute magnitude resulting from the
RR~Lyr calibration with results of studies of Galactic Cepheids. The
Galactic calibrations of Cepheids fall into three categories: based on
Hipparcos direct parallaxes (Feast and Catchpole 1997, Lanoix, Paturel
and Garnier 1999), classical, pre-Hipparcos ones (Laney and Stobie 1994,
Gieren \etal 1998) and based on statistical parallaxes (Luri \etal
1998). They give the brightest,  moderate and faintest luminosity of
Cepheids at a given period and as a consequence the long, classical and
short distance to the LMC. We will not discuss in detail any of these
calibrations here.

As we already mentioned  the Hipparcos parallaxes of Cepheids are very
uncertain and may be biased by many factors. Analysis of the Hipparcos
data by Feast and Catchpole (1997) and  Lanoix \etal (1999) lead to
essentially the same results that the absolute {\it V}-band magnitude of
the Galactic Cepheids with 10-day period  is about $-4.22$~mag. The
classical calibration predicts the mean absolute magnitude of the
Galactic Cepheids of 10-day period equal to $M_V^{C,10}=-4.07$~mag
(Laney and Stobie 1994, Gieren \etal 1998). Finally, the statistical
parallaxes method predicts much fainter Cepheids: $M_V^{C,10}=-3.86$~mag
for 10-day period object (Luri \etal 1998).

Comparing these calibrations with the absolute magnitude inferred from
comparison of Cepheids with RR~Lyr stars in the Magellanic Clouds and
the most likely calibration of RR~Lyr stars we  find  that the Galactic
calibration based on statistical parallaxes (Luri \etal 1998) is closest
to our result. It is worth noting that statistical parallaxes method for
both Cepheids and RR~Lyr stars give consistent results. Calibration of
RR~Lyr of Popowski and Gould (1998) is based, among others, on
statistical parallaxes determination.

\Acknow{We would like to thank Prof.\ Bohdan Paczy\'nski for many
discussions and help at all stages of the OGLE project. We thank  
Drs.\ K.~Z.\ Stanek and D.\ Sasselov for valuable comments on the paper.
The paper was partly supported by  the Polish KBN grants 2P03D00814 to
A.\ Udalski and 2P03D00916 to M.\ Szyma{\'n}ski. Partial support for the
OGLE  project was provided with the NSF grants AST-9530478 and
AST-9820314 to B.~Paczy\'nski.}

\end{document}